\documentclass{ceurart}
\usepackage{listings}
\usepackage[most]{tcolorbox}
\begin{document}

\copyrightyear{2025}
\copyrightclause{Copyright for this paper by its authors.
  Use permitted under Creative Commons License Attribution 4.0
  International (CC BY 4.0).}

\conference{8th Conference on Technology Ethics (TETHICS2025), November 11–12, 2025, Vaasa, Finland}
\title{Can We Trust AI Agents? A Case Study of an LLM-Based Multi-Agent System for Ethical AI}




\author[1]{José Antonio Siqueira de Cerqueira}[%
orcid=0000-0002-8143-1042,
email=jose.siqueiradecerqueira@tuni.fi,
]
\cormark[1]
\address[1]{Tampere University, Tampere, Finland}

\author[2]{Mamia Agbese}[%
orcid=0000-0002-5479-7153,
email=mamia.o.agbese@jyu.fi,
]
\address[2]{University of Jyväskylä, Jyväskylä, Finland}

\author[3]{Rebekah Rousi}[%
orcid=0000-0001-5771-3528,
email=rebekah.rousi@uwasa.fi ,
]
\address[3]{University of Vaasa, Vaasa, Finland}

\author[1]{Nannan Xi}[%
orcid=0000-0002-9424-8116,
email=nannan.xi@tuni.fi,
]

\author[1]{Juho Hamari}[%
orcid=0000-0002-6573-588X,
email=juho.hamari@tuni.fi,
]

\author[1]{Pekka Abrahamsson}[%
orcid=0000-0002-4360-2226,
email=pekka.abrahamsson@tuni.fi,
]

\cortext[1]{Corresponding author.}

\begin{abstract}
AI-based systems, including Large Language Models (LLMs), impact millions by supporting diverse tasks but face issues like misinformation, bias, and misuse. AI ethics is crucial as new technologies and concerns emerge, but objective, practical guidance remains debated. This study explores the extent to which trustworthiness-enhancing techniques in LLMs can support the development of ethically aligned AI software. We adopt a single exploratory cycle of Design Science Research (DSR). First, we identify trustworthiness-enhancing techniques for LLMs: multi-agents, distinct roles, structured communication, and multiple rounds of debate. Second, we design a multi-agent prototype LLM-MAS in which agents address real-world AI ethics issues from the AI Incident Database. Finally, we evaluate the prototype across three case scenarios using thematic analysis, hierarchical clustering, a baseline comparison, and code execution. The system generates approximately 2,000 lines of code per case, compared to only 80 lines in baseline trials. Results reveal terms like bias detection, transparency, accountability, user consent, GDPR compliance, fairness evaluation, and EU AI Act compliance, showing this prototype ability to generate extensive source code and documentation addressing often overlooked AI ethics issues. However, practical challenges in source code integration and dependency management may limit its use by practitioners.
\end{abstract}


\maketitle

\section{Introduction}

Artificial Intelligence (AI) is emerging as a transformative force, reshaping industries, economies, and everyday life. Despite its rapid development and adoption, a number of negative reports on its use highlight the importance of adhering to ethical norms and principles \cite{Vakkuri2021ECCOLAjournal}. Several ethical guidelines are available with plenty of ethical principles providing high level and abstract guidance for developers and stakeholders on AI ethics \cite{Cerqueira2022Guide}. These are important, but there is a lack of practical guidance for developers to operationalise ethics in AI \cite{Cerqueira2022Guide}. As new AI-based technologies emerge, ethics in AI will also become increasingly critical, such as the latest breakthrough: Large Language Models (LLMs).

LLMs are becoming ubiquitous and have significant impact on decision-making processes and human interactions \cite{liu2023trustworthy,Ray2023ChatGPTReview}. LLMs are advanced AI algorithms capable of generating, interpreting, and predicting text based on vast amounts of data they have been trained on \cite{chang2023survey}. Of these LLMs, Generative Pre-trained Transformer (GPT) LLMs, such as GPT-4o, have showcased unprecedented proficiency in the human language, coding, logic, reasoning, and other associated natural language tasks \cite{Ray2023ChatGPTReview}. They excel at guiding complex conversations and are used in countless endeavours such as software engineering (SE) - as in AI for Software Engineering (AI4SE) \cite{DBLP:conf/fose-ws/Lo23} - and qualitative research \cite{ni2024mechagents}.

Within the AI4SE field, the capabilities of LLMs are being explored in the software development, maintenance, and evolution \cite{ozkaya2023application,peng2023software,ni2024mechagents,DBLP:conf/fose-ws/Lo23}, namely LLM4SE. Accordingly, they find applications across various stages of the software development life cycle, including requirement analysis, software design, code implementation, testing, refactoring, defect detection, and maintenance \cite{peng2023software}. Albeit the notable use of LLM in SE, to the best of our knowledge there are no studies that focus on the use of LLM in the ethical AI development. In qualitative research, LLMs are also gaining prominence, particularly as a supplement to tasks traditionally performed by humans, e.g., analysing qualitative data \cite{hamilton2023exploring,rasheed2024can,David2023ExploringTheUseofAIQualitative}. Specifically, in the coding process - where qualitative data is organised and interpreted by assigning codes or labels to text segments or different data forms - LLMs have demonstrated significant utility \cite{siiman2023opportunities,rasheed2024can}.

However, several concerns arise, especially related to trustworthiness, as more practitioners are relying on LLMs to perform their task \cite{DBLP:conf/fose-ws/Lo23}. Several studies are drawing attention to possible problems in adopting LLMs for SE, in particular when syntactically correct but non-functional code is produced, which affects the reliability and effectiveness of LLM-based code generation \cite{Hou2023LLM4SEaSLR}. Pearce et al. \cite{Pearce2022Asleep} used GitHub Copilot to produce 1,689 programs, and found that 40\% of them have security vulnerabilities. Liu et al. \cite{Liu2024RefiningChatGPTCode} systematically analysed ChatGPT code generation reliability identifying quality issues as many of the programs generated provided wrong output or contained compilation or runtime error.

Effective AI4SE should be trustworthy and synergistic with the practitioner's workflow, otherwise ``such AI4SE solutions risk becoming obstacles rather than facilitators" \cite{DBLP:conf/fose-ws/Lo23}. Currently, there is a growing need for context-dependent empirical studies that explore how trust in LLMs affects their adoption in software engineering tasks \cite{DBLP:conf/fose-ws/Lo23}. Furthermore, there remains a significant gap in research focused on the practical operationalisation of AI ethics, particularly through the application of LLMs. To date, no studies have attempted to explore the development of ethical AI-based systems using LLMs. We aim to explore trustworthiness-enhancing techniques in LLMs in the ethical AI development context. 

The following Research Question guides this study: 

\begin{itemize}
    \item RQ: To what extent can trustworthiness-enhancing techniques in LLM-based systems support the development of ethically aligned AI software?
\end{itemize} 

To this end, we employed an adapted single cycle of the Design Science Research (DSR) method \cite{hevner2004design} to identify trustworthiness-enhancing techniques, build a prototype, evaluate it through three case studies and communicate the results. We identified four trustworthiness-enhancing techniques for LLM-based systems in the literature and implemented them: \textbf{(1) multi-agent collaboration, (2) specialised roles, (3) structured communication, and (4) multiple rounds of debate.} Furthermore, we evaluated the prototype through three case studies based on real-world AI incidents in recruitment, deepfake detection, and image classification. Each case is analyzed using (1) LLM for thematic analysis, (2) iRaMuTeQ for hierarchical clustering, (3) a comparative baseline with standard ChatGPT GPT-4o prompts, and (4) source code execution to verify whether the generated code runs, without assessing its runtime behaviour or correctness in depth. \textbf{The prototype can generate around 2,000 lines of code (with proper documentation), encompassing themes such as bias detection, transparency, accountability, user consent and GDPR compliance, fairness evaluation and compliance with the EU AI Act.} The hierarchical clustering analysis revealed the ability of the prototype to generate outputs related to legal aspects of ethics in AI through words such as: bias\_type, evaluate\_bias, bias\_score. The baseline study produced around 80 lines of output without source code while covering considerably less AI ethical issues. Therefore, the evaluations conducted suggest that the trustworthiness-enhancing techniques employed can be beneficial in the AI ethical development context. However, practical challenges in source code integration and dependency management may limit its use by practitioners, e.g., deprecated dependencies and modularity. We hope to shed light both on how to improve trustworthiness in LLM and on how to help practitioners to operationalise ethical principles in the development of AI-based systems.

\section{Background and Related Work}

\subsection{Large Language Models}
LLMs have been pre-trained on vast amounts of text data, often scraped from the internet, allowing them to learn patterns and structures inherent in human language \cite{hoffmann2022training}. One of the key features of LLMs is their ability to generate contextually relevant and apparently coherent text across a wide range of topics \cite{Ray2023ChatGPTReview}. These models can perform tasks such as language translation, text summarization, question answering, and creative writing \cite{hoffmann2022training}. They achieve this by leveraging the vast amounts of data they have been trained on to predict the next word or sequence of words in each context \cite{floridi2020gpt3}. Some well-known examples of LLMs include Generative Pre-trained Transformer (GPT) models developed by OpenAI \cite{openai2023chatgpt}, particularly GPT-4o and o3.

\subsection{Trustworthiness in Large Language Models}
\label{trustworthiness}

Recent advancements and use of LLMs have raised concerns about their ethical implications \cite{Wang2023DecodingTrustAC,Liang2023HolisticHELM}. Issues like information hallucination, biases, and the need for factual correctness are critical in determining the trustworthiness of these systems in real-world applications \cite{Lemon2022,Du2023ImprovingFA}. Trustworthiness refers to the quality or attribute of being reliable, dependable, and deserving of trust. It encompasses several key components that establish trust in a person, organisation, product, or system \cite{sun2024trustllm}. In the context of LLMs, trustworthiness largely refers to their reliability and ethical adherence to social norms \cite{liu2023trustworthy,sun2024trustllm}. Notably, several studies have formulated taxonomies accompanied by evaluation methodologies, wherein the taxonomy acts as a measure for assessing trustworthiness, with the identified aspects serving as evaluation parameters. Among these efforts, the Holistic Evaluation of Language Models (HELM) devised by Liang et al. \cite{Liang2023HolisticHELM} introduces a comprehensive taxonomy consisting of seven \textbf{metrics} — accuracy, calibration, robustness, fairness, bias, toxicity, and efficiency — and applied this framework to analyze 30 different models. Notably, the authors stress the significance of adopting a multi-metric approach, enabling metrics prioritization beyond mere accuracy.

Wang et al. \cite{Wang2023DecodingTrustAC} provide eight trustworthiness \textbf{perspectives} of language models: toxicity, stereotype and bias, adversarial robustness, out-of-distribution robustness, privacy, robustness to adversarial demonstrations, machine ethics, lastly, fairness. The authors state that they only provided objective definitions for some of the perspectives coined - due to their intrinsic subjectivity, e.g., fairness toxicity - leaving the subjective exploration of model behaviours based on human understanding as future work. As scholars delve into characterizing trustworthiness within LLMs, a unified terminology for measuring it remains unclear. Furthermore, it is seen that there is no clear consensus on a framework to build aligned LLMs, nor clear assessment guidelines, both posing as unresolved issues \cite{liu2023trustworthy}.

\subsection{Trustworthiness enhancing techniques in LLM4SE}
\label{techniques}

Hong et al. \cite{Hong2023MetaGPT} propose MetaGPT to simulate a group of agents as software company. Worth noting the use of \textbf{specialised roles} (Product Manager, Architect, Project Manager, Engineer, and QA Engineer), \textbf{workflow for the agents} (all agents work in a sequential fashion), and a \textbf{structured communication interface}. An interesting point regarding the specialised roles, is the ability for agents to have skills, i.e., perform actions, such as being able to run code and search on the web. The authors argue that the incorporation of aforementioned techniques can considerably enhance code generation, as well as mitigating hallucinations risks.

Qian et al. \cite{Qian2023CommunicativeChatDev} developed ChatDev, a chat-based software development framework using LLMs to enhance communication and collaboration across various roles, aimed at reducing hallucination risks like bugs and missing dependencies. Though, it does generate different outputs for the same inputs and carries risks due to untested code. 

Meanwhile, Wu et al. \cite{Wu2023AutoGenEN} from Microsoft introduced AutoGen, a multi-agent LLM framework enhancing coding productivity by structuring tasks among specialised roles (Commander, Writer, Safeguard) that interact to refine outputs. Through an ablation study they demonstrated that this method significantly outperforms single-agent systems by breaking down complex tasks into manageable components. 

The aforementioned studies run in line with a joint study by MIT and Microsoft, Du et al. \cite{Du2023ImprovingFA}. Albeit not using LLM-based multi-agents to directly generate code, the authors argue that using multiple agents along with multiple rounds of debate improve the reasoning and factuality of the generated solution compared to not using them. Hence, multiple agents and rounds of debate can reduce hallucinations and increase the overall trustworthiness of using LLMs. Such techniques can exhibit emergent behaviours that are not evident when considering individual agents in isolation \cite{Hong2023MetaGPT,Talebirad2023MultiAgentCH}.

\subsection{AI Ethics}

Despite the recent academic and industry interest, AI ethics is a long-standing area of research, in which diverse incidents have recently raised public concerns about its use and development \cite{vakkuri2024MakingEthicsPractical}. Consequently, in the past few years a variety of principles and guidelines have emerged, from various sources (academia, industry, civil society) to frame and define what is ethical AI \cite{jobin2019globallandscape}. Ryan and Stahl \cite{Ryan2020ArtificialIE} provide 11 ethical principles along with ethical issues related to AI: 1) Transparency, 2) Justice and Fairness, 3) Non-maleficence, 4) Responsibility, 5) Privacy, 6) Beneficence, 7) Freedom and Autonomy, 8) Trust, 9) Sustainability, 10) Dignity, 11) Solidarity. These ethical principles serve as the basis of our analysis.

The European Parliament is currently putting into force the world's first regulation on AI \cite{europaActFirst_EUAIAct}, which highlights the fact that the plethora of guidelines available are in fact soft law, with no legally binding nor real consequences \cite{Cerqueira2022Guide}. Nevertheless, it is reminiscent of the abstract principles on how AI ethics is mainly approached \cite{vakkuri2024MakingEthicsPractical,jobin2019globallandscape,Correa2023WorldwideAIethics200}. Thus, the obstacles in operationalising ethical principles in AI-based systems stem from the subjectivity necessary to interpret and put into practice highly abstract principles by the practitioners \cite{Cerqueira2022Guide}. Despite its importance, it has often been a task commonly taken as an afterthought \cite{Vakkuri2021ECCOLAjournal}. As opposed to other efforts available in the literature, we aim to study the use of LLM-based multi-agent systems in the development process of AI-based systems, taking into consideration the operationalization of ethical principles from the first stage of the development process onwards. The use of LLM agents in developing ethically-aligned AI systems introduces significant complexity and novelty, making their ethical implications a compelling area of study. LLMs face unique challenges in producing accurate, unbiased text and source code while adhering to ethical guidelines. Their dynamic and often unpredictable outputs in multi-agent interactions provide a valuable opportunity to study the application of ethical principles in real-world settings, although there is a noted lack of literature on the operationalization of AI ethics through LLM agents.

\subsection{LLMs for Qualitative Analysis}

Qualitative analysis is another opportune expanding area where LLMs can be applied \cite{dai2023llmforthematicanalysis,Drpal2023,David2023ExploringTheUseofAIQualitative}, particularly in qualitative research within the field of software engineering \cite{bano2024large,rasheed2024can}. Thematic analysis is labour-intensive and time-consuming for humans, and are prone to mistakes and bias \cite{Braun2006}. Therefore, LLMs present an opportunity to improve the accuracy and efficiency of this method. In thematic analysis, coders require multiple rounds of discussion to achieve consensus and resolve ambiguities for an in-depth understanding of the data \cite{Braun2006}. 

Thus, LLMs can tackle the challenges of traditional qualitative research, such as the time-intensive nature of data analysis, limited generalisability, consistency issues, and personal subjective biases \cite{bano2024large,hamilton2023exploring}. Hamilton et al. \cite{hamilton2023exploring} argue that both humans and LLMs have identified unique and overlapping themes from interview data, indicating the potential for recognising patterns and themes in qualitative data. Drápal et al. \cite{Drpal2023} proposed a framework for collaboration between legal experts and LLM for performing thematic analysis. The LLM used could identify themes that would probably be missed due to human error. Also, they discuss that LLM can perform the initial coding of the data with reasonable quality. De Paoli \cite{DePaoli2023} states that ``there is no denying that the model can infer codes and themes and even identify patterns which researchers did not consider relevant and contribute to better qualitative analysis.''

Differently from the aforementioned approaches, our study aims to provide case studies using LLMs automating the whole thematic analysis process, as well as implementing the trustworthiness-enhancing techniques found to propose a prototype that generates source code in the AI ethical development context. For the best of our knowledge, there are no works in the literature that aims to use LLM to operationalise AI ethical principles.

\section{Methodology}

In this study, we used an adapted DSR approach \cite{hevner2004design} to explore and enhance the trustworthiness of LLMs for software engineering tasks. We chose this method because it is a problem-solving paradigm that advances knowledge through the building and evaluation of artefacts \cite{hevner2004design}. In our study, its role is to provide a structured approach in four phases: (1) exploration, (2) prototyping, (3) evaluation through case studies and (4) communication of results. Due to the limited scope of this initial case study, we implemented only a single exploratory cycle. The evaluation phase consists of applying the LLM-MAS prototype to three case scenarios, each derived from real-world AI incidents, and analysing the generated outcomes using thematic analysis, hierarchical clustering, comparative baselines, and source code execution. These case studies serve to examine how the identified trustworthiness-enhancing techniques impact the development of ethically aligned AI-based systems in practice. This Section will cover phases 1 to 3, while phase 4 will be presented in Section \ref{results}.

\subsection{Exploring}

Initially, we explored the existing literature and identified the need to improve trustworthiness in AI4SE, as presented in Section \ref{trustworthiness}. This need is most clearly demonstrated in the work of David Lo \cite{DBLP:conf/fose-ws/Lo23}. Furthermore, we identified recently proposed techniques in the literature to enhance trustworthiness in LLMs, particularly in the context of software engineering (i.e. LLM4SE). In summary, the trustworthiness-enhancing techniques identified and presented in Section \ref{techniques} provide the basis for our conceptual approach, which is visualised in Table \ref{tab:techniques}.

\begin{table}[ht]
\centering
\caption{Trustworthiness-enhancing techniques adopted}
\label{tab:techniques}
\begin{tabular}{|l|l|}
\hline
\textbf{Trustworthiness-enhancing technique} & \textbf{Description} \\
\hline
Multi-agent collaboration & Multiple agents interact on a task \\
\hline
Specialised roles & Agents assume roles, e.g., Developer or Ethicist \\
\hline
Structured communication & Predefined message passing structure \\
\hline
Multiple debate rounds & Iterative dialogue to refine results \\
\hline
\end{tabular}
\end{table}

Next, we will present a detailed overview of the technical implementation of the prototype.

\subsection{Prototype: LLM-MAS Technical Implementation}

We devised a prototype of an LLM-based Multi-Agent System (henceforth referred to as the LLM-MAS). The LLM-MAS is a Python script consisting of three agents that make use of the GPT-4o-mini model via the OpenAI API. The model was selected on the basis of its cost-efficiency, accessibility, and strong reasoning performance, which rendered it a practical choice for iterative experimentation involving multiple agents. Rather than being equipped with memory or tool access, the agents are LLMs with custom instructions in the form of system prompts that describe their roles and the communication structure. Consequently, they have different specialised roles, and their conversations are structured in the same way. Specifically, there are \textbf{two senior Python developers and one AI ethicist}. The \textbf{conversation structure} comprises: \textbf{reply, reflection, code, and critique}. The agent prompts are available in the \href{https://zenodo.org/records/15425970}{Zenodo}: agent1\_prompt\_PythonDev.txt, agent2\_prompt\_PythonDev.txt and agent3\_prompt\_AIEthicist.txt.

\begin{figure}[htbp]
\centering
\includegraphics[width=0.37\textwidth]{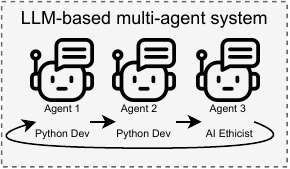}
\caption{Prototype overview}
\label{fig:LLM-based_Multi-Agents}
\end{figure}

\textbf{The workflow for the agents is sequential}. A round is defined as a sequence of conversations between agents 1, 2 and 3. \textbf{The conversation history is appended after each agent output, and each agent receives the entire conversation history as an input.} The number of rounds is modifiable. We selected five rounds per project to allow agents enough interaction to develop and refine ideas collaboratively, while keeping the output concise and manageable for analysis. Figure \ref{fig:LLM-based_Multi-Agents} presents an overview of our prototype.

The agents engage in a structured conversation, through multiple rounds, to discuss and collaborate on a project. That is, as a team, they receive an input -- a project description (PD) -- that they should work on collaboratively to implement, through structured, iterative discussions. 

\subsection{Evaluation: Case Studies}

We begin with an overview of the evaluation approach, followed by an in-depth explanation of each component.

\begin{figure}[htbp]
\centering
\includegraphics[width=0.67\textwidth]{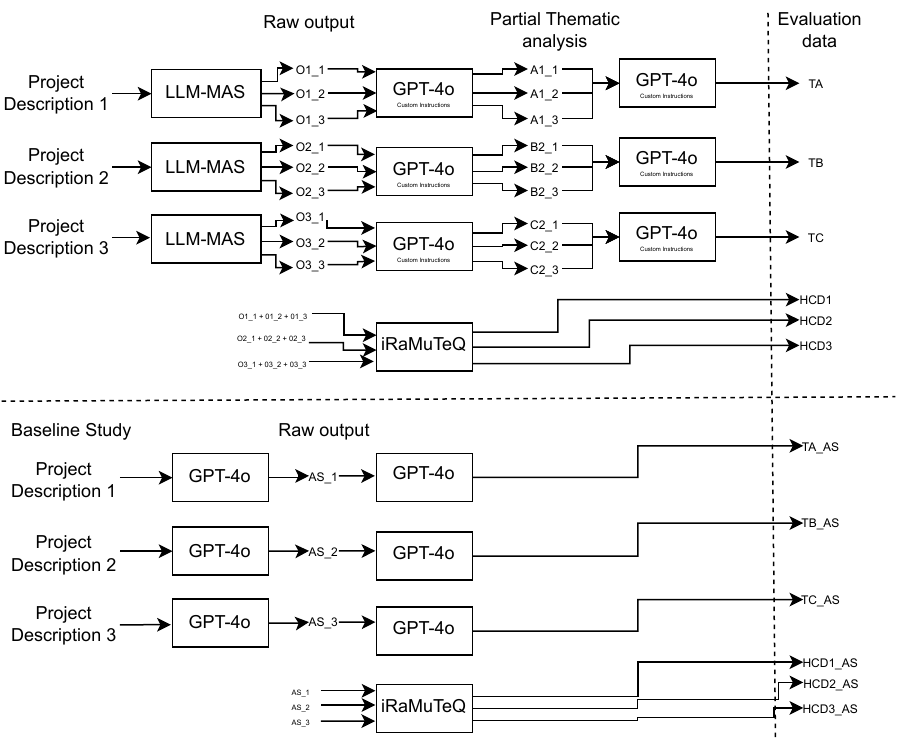}
\caption{Evaluation overview}
\label{fig:EvaluationOverview}
\end{figure}

To evaluate the effectiveness of the prototype, we conducted a case study-based evaluation across three distinct scenarios inspired by real-world AI incidents. \textbf{Each case study} involved applying the LLM-MAS prototype to a different project description, enabling us to assess how the system behaved in various ethical and technical contexts. For each case, we employed a multi-faceted evaluation strategy. First, we performed a \textbf{thematic analysis} using an LLM to automatically identify and categorise key ethical and technical themes in the output. This helped us to assess the system’s ability to address ethical requirements. Secondly, \textbf{hierarchical clustering} (via iRaMuTeQ) was used to group related text segments, providing an additional layer of validation by highlighting patterns in the generated content. Thirdly, we conducted a \textbf{comparative baseline study}, prompting the same project descriptions directly to GPT-4o in ChatGPT interface (without any multi-agent configuration), to evaluate the added value of the proposed trustworthiness-enhancing techniques. Figure \ref{fig:EvaluationOverview} presents an overview of the evaluation methodology approached.

Due to the probabilistic nature of LLMs, running the same input multiple times can yield different outputs. This means that even when using the same system configuration, each run may produce a different conversation, source code, and thematic analysis. Prior work has similarly observed this variability in multi-agent LLM systems \cite{Qian2023CommunicativeChatDev,DBLP:conf/fose-ws/Lo23}. Thus, in order to enrich our analysis, we ran the system three different times to obtain distinct output (yielding the raw output, e.g., $O1_1$, $O1_2$, $O1_3$ for the first project description).

Then, the collected data is analysed through the use of ChatGPT GPT-4o with a specific \textit{custom instruction} acting as qualitative analyst, available in the file Custom\_instructions\_Qualitative\_GPT.txt in \href{https://zenodo.org/records/15425970}{Zenodo}. The prompt used as a header for the thematic analysis in available in file Prompt\_Perform\_thematic\_analysis.txt. Each of the raw output produces a partial thematic analysis (e.g., $B2_1$, $B2_2$, $B2_3$ for the second project description). Consequently, the partial thematic analysis for each project description is summarised into one thematic analysis through the use of the same GPT-4o, yielding $TA$, $TB$ and $TC$. The prompt serving as header for the merging of the thematic analysis is available in file Prompt\_Merging\_Thematic\_Analysis.txt, equipped with the same custom instruction.

We used Descending Hierarchical Classification (DHC) from iRaMuTeQ v0.7-alpha running in R v4.2. The input for the generation of each hierarchical clustering is the append of the raw output of each project description (e.g., $O3_1$, $O3_2$, $O3_3$ together generates the hierarchical clustering for the third project description.

With the aim of understanding the benefits of this LLM-MAS approach, a baseline study is also performed, in which a project description is prompted to a regular ChatGPT-4o using OpenAI user interface, without any of the techniques found. For example, for project description 1, we have only one output ($AS_1$), that serves as input for creation of its thematic analysis also with ChatGPT-4o, and for the iRaMuTeQ analysis, yielding $TA_AS$ and $HCD_AS$ respectively.

Finally, we performed a practical evaluation of the prototype by running the source codes generated by the LLM. This step is for evaluating the system's practical utility in real-world software development scenarios. Our methodology offers an initial framework for assessing the trustworthiness and reliability of LLMs in the software engineering domain. The results and discussions arising from the comparison between the different thematic analysis and hierarchical clustering, coupled with the baseline study and source code execution are presented next.

\section{Results}
\label{results}
In this Section, we present the results for each of the project descriptions used in the case study. Each project description is derived from real AI incidents available at \href{https://incidentdatabase.ai/}{https://incidentdatabase.ai/}. The selected incidents focus on AI applications in recruitment, deepfake detection, and image classification. They were selected due to their high relevance to AI ethics and the need for compliance with regulatory standards. All data generated during the study -- including agent prompts, custom instructions, and thematic analyses -- are available in \href{https://zenodo.org/records/15425970}{Zenodo}.

\subsection{Project Description 1}

The first project description derived from \href{https://incidentdatabase.ai/cite/37#r620}{Incident 37: Female Applicants Down-Ranked by Amazon Recruiting Tool}: 

\begin{tcolorbox}[colback=gray!10]
Develop an AI-powered recruitment tool designed to screen resumes impartially, complying with the EU AI Act. The project aims to eliminate biases related to gender and language, ensuring fair evaluation of all applicants. The AI Ethics Specialist will guide the team in addressing ethical concerns and risk levels. The senior Python developers will utilise NLP to process resumes, referencing relevant EU AI Act guidelines. 
\end{tcolorbox}

It is important to note that the project description references the EU AI Act, but the LLM used may not have had access to its full content at the time this study was conducted. In future work, we will introduce this document as a vector to be embedded and accessed by an open-source LLM. Moreover, according to David Lo \cite{DBLP:conf/fose-ws/Lo23}, guaranteeing that AI4SE solution will comply with legislation is one way to enhance trustworthiness in AI4SE, since practitioners need to avoid conflict with laws.

\begin{figure}[htbp]
\centering
\includegraphics[width=0.57\textwidth]{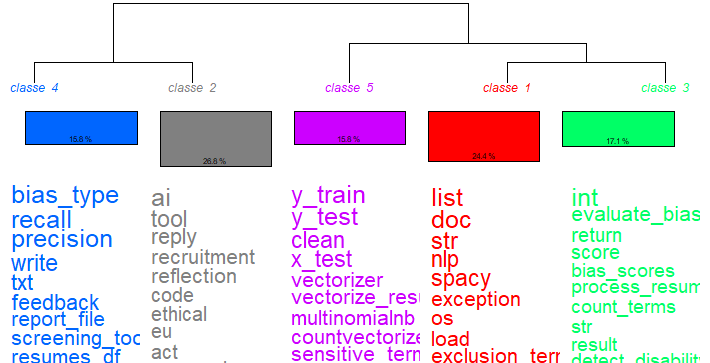}
\caption{Hierarchical clustering dendogram - PD1}
\label{fig:CHD1}
\end{figure}

Raw data obtained vary from 1,712 lines to 1,976 lines of discussion plus source code. The thematic analysis in Table \ref{tab1} was obtained from the merge of the three thematic analyses attained from each output. In Figure \ref{fig:CHD1}, 5 classes were devised, however, classes such as 1, 3, 4 and 5 are related to technical implementation. Nevertheless, it is possible to understand that themes 1, 2 and 3 are closely related to the second class, where ethical, EU, act appear. Moreover with class number 3, w.r.t functions as \texttt{evaluate\_bias} and \texttt{bias\_score}.

In all cases, the source codes produced were spread around the text. As mentioned, this can be almost 2,000 lines long. This makes the source code difficult to differentiate, and sometimes there are pieces that need to be glued together from different rounds of conversation. After trying to run the source code available in O1\_1, it was found that some packages were needed to be installed manually. Furthermore, deprecated packages were being used: \texttt{sklearn} PyPI package is deprecated, use \texttt{scikit-learn}. This results in extra effort by the practitioners, impacting on their workflow.

\begin{table}
\caption{Thematic Analysis PD 1}
\label{tab1}
\begin{tabular}{|l|p{7cm}|} 
\hline
\textbf{Theme} & \textbf{Codes} \\
\hline
Ethical and Regulatory AI Development & Compliance with the EU AI Act, Ethical considerations in AI development, Bias elimination strategies \\
\hline
Fair and Transparent AI Implementation & Technical development of the AI tool, Integration of bias detection mechanisms, ensuring fairness in AI operations \\
\hline
Iterative Model Training and User-Centered Validation & Scenario-based testing, Continuous model refinement, Incorporating actionable user feedback \\
\hline
Transparency in AI Processes & Documentation of AI processes, Clear communication of updates and improvements, Stakeholder engagement and transparency \\
\hline
\end{tabular}
\end{table}

In regard to the baseline study, only 78 lines of text were obtained, while no code was produced. On the other hand, three themes were devised from the thematic analysis: Ethical AI Development, Technical Implementation and Fairness, and Compliance. It was not possible to generate the hierarchical clustering for any of the baseline studies due to small size of text data available. The quality between both studies is very disparate.

\subsection{Project Description 2}

The second project description derived from \href{https://incidentdatabase.ai/cite/635/#r3637}{Incident 635: AI-Generated Fake News Targets Black Celebrities on YouTube}: 

\begin{tcolorbox}[colback=gray!10]
Create a deepfake content authentication system that uses AI to verify the authenticity of video, image and audio content. This system will distinguish between real and manipulated media, reducing the spread of harmful deepfake content. The AI Ethics Specialist will ensure the tool adheres to ethical standards. The senior Python developers will enhance the system’s detection capabilities.
\end{tcolorbox}

In this PD, there is no reference to EU AI Act directly. However, it is worth pinpointing that it is available in the prompt of the third agent, the AI Ethicist. Raw data obtained vary from 1414 lines to 2178 lines of text data. The discussion here progresses in a similar way to the previous one. 

\begin{figure}[htbp]
\centering
\includegraphics[width=0.57\textwidth]{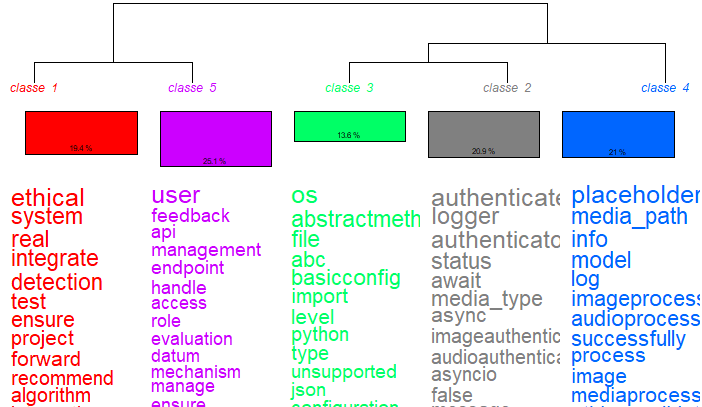}
\caption{Hierarchical clustering dendogram - PD2}
\label{fig:CHD2}
\end{figure}

As the relationship between the thematic analysis in Table \ref{tab2} and the hierarchical clustering produced in Figure \ref{fig:CHD2}, it is possible to highlight the class number 1 with theme number 2, Ethical Compliance and AI Responsibility. Moreover, class number 2 is closely related to theme Security and Collaboration, with words such as authentication and logger. Similarly to the previous PD, no source code was produced in the baseline study. This yielded 78 lines of text and themes such as Technological Framework, Ethics and Compliance, User Experience and Support, Strategic Objective. While the study with the proposed prototype provides a more fruitful outcome, yielding almost 2,000 lines of text, the baseline study provides a generic text that can only serve as study material.

In relation to the source code produced, on top of the overhead to developers encountered in the previous PD, in O2\_2 it is perceived that our prototype created modules, as in \texttt{from your\_module import VideoAuthenticator, ImageAuthenticator, AudioAuthenticator}. Again, this poses a problem to practitioners when working with this tool, due to the fact that they will need to go through all the text and manually search for the source code that he/she needs to copy and paste somewhere else, test it, install dependencies, and manage modules.

\begin{table}
\caption{Thematic Analysis PD 2}\label{tab2}
\begin{tabular}{|l|p{7cm}|} 
\hline
\textbf{Theme} & \textbf{Codes} \\
\hline
System Architecture and Design & Modularity and Scalability, AI Integration and Media Processing, Role-Based Access Control (RBAC) \\
\hline
Ethical Compliance and AI Responsibility & AI Ethics and Bias Detection, Transparency and Accountability, User Consent and GDPR Compliance \\
\hline
Security and Collaboration & User Authentication and Data Security, Role Management and Access Control, Collaboration between Developers and Ethics Specialists \\
\hline
System Reliability and Continuous Improvement & Error Handling and Logging, Feedback Mechanisms, Testing and Maintenance \\
\hline
\end{tabular}
\end{table}

\subsection{Project Description 3}

The third project description derived from \href{https://incidentdatabase.ai/cite/16#r88}{Incident 16: Images of Black People Labeled as Gorillas}: 

\begin{tcolorbox}[colback=gray!10]
Develop an AI-based image processing and classification system that complies with the EU AI Act, ensuring accurate and respectful labelling of images. The system must avoid racial biases and harmful misclassifications. The AI Ethics Specialist will focus on identifying and mitigating psychological harm and harm to social systems. The senior Python developers will implement and refine image classification algorithms.
\end{tcolorbox}

\begin{figure}[htbp]
\centering
\includegraphics[width=0.57\textwidth]{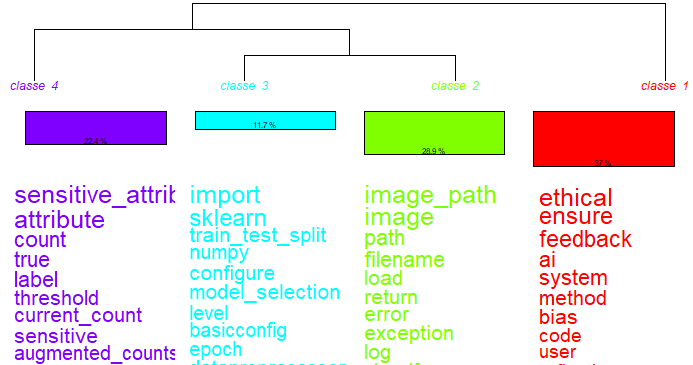}
\caption{Hierarchical clustering dendogram - PD3}
\label{fig:CHD3}
\end{figure}

This discussion follows in a similar fashion to the previous discussions. Raw data obtained vary from 1,676 lines to 2,122 lines of text data. The thematic analysis in Table \ref{tab3} revolves around Ethical AI Development and Compliance and Technical Implementation, Optimization and Robustness. Class number 1 is related to theme number 1, however, the rest of the classes available contain words that are related to technical implementations, such as \texttt{image\_path}, \texttt{import}, \texttt{sklearn}, \texttt{sensitive\_attribute}, which are related to theme number 2. Interesting to note that \texttt{sklearn} is present in a class in Figure \ref{fig:CHD3}, even though it had already been recognised to be deprecated. This highlights the challenge of relying on static LLM knowledge in a rapidly evolving software landscape, where tools and dependencies can quickly become outdated.

\begin{table}
\caption{Thematic Analysis PD 3}\label{tab3}
\begin{tabular}{|l|p{7cm}|} 
\hline
\textbf{Theme} & \textbf{Codes} \\
\hline
Ethical AI Development and Compliance & Ethical AI, Compliance, Bias Detection and Mitigation, Fairness Evaluation, Transparency and Accountability \\
\hline
Technical Implementation, Optimization, and Robustness & Data Management, Model Design and Implementation, Long-term Monitoring and Continuous Improvement, Error Handling and Modular Design \\
\hline
Stakeholder Engagement and Continuous Improvement & Stakeholder Engagement, User Feedback Integration, Collaboration, Transparency in Operations \\
\hline
Data-Driven Evaluation & Statistical Analysis, Confidence Scores, Class-wise Analysis \\
\hline
\end{tabular}
\end{table}

\section{Discussion}
\label{Discussion}

The results suggest that the trustworthiness-enhancing techniques employed in the prototype, such as multi-agent collaboration, specialised roles, structured communication, and multiple debate rounds, can improve the quality of generated artefacts, as observed in \cite{Hong2023MetaGPT,Qian2023CommunicativeChatDev,Wu2023AutoGenEN}. While LLM agents are increasingly integrated into traditional software engineering tasks (e.g., requirements engineering, code generation, software QA, and maintenance) \cite{He2025,liu2024largelanguagemodelbasedagents}, none systematically study their use in the operationalisation of AI ethics. Thus, our results extend prior work by adding the ethical AI dimension. Specifically, we implement LLM-trustworthiness techniques not only to complete software engineering tasks, but to tackle AI ethical issues, that is, to bring fairness, accountability, transparency, privacy, and legal alignment (EU AI Act, GDPR) into the generated output from the start.

The prototype produced more code and richer ethics-related artefacts (e.g. bias-related functions, consent handling, documentation) compared to the baseline study. This is consistent with evidence that multi-agent debate improves reasoning and factuality \cite{Du2023ImprovingFA} and suggests that improving role specialisation in LLM-based agents is a next step \cite{He2025}. Nonetheless, we observed practical issues that matter for adoption: scattered modules, dependency management, and extra packaging/integration work. These align with calls to improve developer–AI synergy in AI4SE \cite{DBLP:conf/fose-ws/Lo23}. For practice, this gives developers tools to consider ethical impacts during AI system development. For research, it invites further work to test, compare, and improve these techniques.

\section{Threats to Validity}

This study has several limitations that may affect the validity and generalisability of its findings. First, the prototype relies on a proprietary LLM by OpenAI, which may evolve over time in undocumented ways. This creates a reproducibility challenge, as future versions of the model may behave differently, even under identical conditions. Secondly, the evaluation relies on an LLM to perform a thematic analysis of the output, which could raise questions regarding its validity and objectivity. Although recent studies (e.g. \cite{hamilton2023exploring}; \cite{Drpal2023}) have demonstrated the potential of LLMs to support, and even enhance, qualitative analysis by identifying patterns that human coders might overlook, this approach remains methodologically debated. In our study, we did not conduct a parallel human-led thematic analysis, which limits our ability to triangulate or verify the consistency of the themes generated. This increases the risk that the LLM-based analysis will reflect the model's own limitations or biases. Future work should either involve expert human analysts to validate the AI-generated codes and themes or adopt a hybrid human-in-the-loop approach to enhance analytical rigour. Third, the thematic analysis and hierarchical clustering processes were fully automated and lacked human validation. While these methods offer efficiency and consistency, they may miss subtle, context-sensitive insights that would be more easily detected by domain experts, particularly in the ethical dimensions of software engineering.

Fourth, although the prototype generates source code, our execution tests were limited to confirming whether the code runs, rather than assessing its functionality, correctness, or ethical alignment. This leaves open questions about the practical reliability and appropriateness of the generated software. Finally, the study was limited to three case scenarios. While these were selected for their ethical relevance, the small sample size restricts the scope of generalisability. Future work should expand on this with broader case coverage, incorporate human-in-the-loop validation, and include a more rigorous assessment of the ethical dimensions of the generated artefacts. It is also important to note that the evaluation was conducted using a case study approach. Our objective was not to produce generalisable results, but rather to explore and gain in-depth insights into how trustworthiness-enhancing techniques in LLM-based systems can support the development of ethically aligned AI. Case studies are particularly valuable for examining complex, real-world phenomena in context, and for uncovering practical challenges and opportunities that may not emerge in more controlled or large-scale experiments \cite{Runeson2012casestudy}.

\section{Conclusion}

This study aimed to explore trustworthiness in AI4SE, specifically in the use of LLM for software engineering (LLM4SE), in the context of ethically aligned AI-based systems development. To the best of our knowledge, there are no studies in the literature that address practical AI ethics using LLM. Our Research Question is: To what extent can trustworthiness-enhancing techniques in LLM-based systems support the development of ethically aligned AI software. In order to provide an answer, we used the Design Science Research method, we identified the motivation and different trustworthiness-enhancing techniques in LLM4SE, proposed a prototype - an LLM-based multi-agent system (LLM-MAS) -, and evaluated it through three case studies. The techniques discovered to improve trustworthiness are (1) multi-agent collaboration, (2) specialised roles, (3) structured communication, and (4) multiple rounds of debate. Each case study had a project description derived from real AI incidents extracted from the AI Incident Database in recruitment, deepfake detection, and image classification. For each project description, LLM-MAS produced about 2,000 lines of text, encompassing source code and documentation to implement the AI-based systems.

Some of the codes produced through thematic analysis include: bias detection, transparency and accountability, user consent and GDPR compliance, fairness evaluation and compliance with the EU AI Act. In relation to hierarchical clustering, some of the words that depict ethics in AI include: ethical, bias, \texttt{bias\_type}, \texttt{evaluate\_bias}, \texttt{bias\_score}. This highlights the ability of LLM-MAS to generate outputs related to legal aspects of ethics in AI (i.e. GDPR and EU AI Act), in addition to theoretical and practical ethical issues, both in the documentation and in the source code, through codes in the thematic analysis (e.g., transparency, fairness, bias) and functions detected in the hierarchical clustering (e.g., \texttt{evaluate\_bias}), respectively. Regarding the baseline study, for each project description it produced around 80 lines of output without source code, covering considerably less AI ethical issues. Therefore, this study underlines the potential of employing trustworthiness-enhancing techniques in LLMs to the development of more ethically aligned AI-based systems.

Nevertheless, there are some issues that hinder the synergy for practitioners \cite{DBLP:conf/fose-ws/Lo23}. Worth highlighting is the lack of practicality, from scraping the source code from the text - even more difficult when it comes to modules -, and installing packages and dealing with deprecated dependencies stemming from the LLM training date. While it is possible to improve trustworthiness and quality in the use of LLM4SE, there are still open issues to work on to make it more practical for practitioners. Whilst this experiment was conducted relying only on the internal knowledge of the Open AI LLM, it is important to note that providing the AI ethical agent with legislation (e.g., GDPR and EU AI Act) or AI ethical methods or guidelines (e.g., ECCOLA \cite{Vakkuri2021ECCOLAjournal}) is a prospective future work that we intend to pursue. Moreover, in-depth analysis of the source code provided by the prototype is necessary to assess ethical alignment. With this study, we hope to further the understanding of the benefits of trustworthiness-enhancing techniques in LLM for the development of ethical AI-based systems. We aim to open source the prototype source code in further studies.


\begin{acknowledgments}
This research was supported by CONVERGENCE of Humans and Machines (220025) and the EVIL-AI “The identification and the mitigation of the negative effects of Artificial Intelligence Agents'' (JAES/2024/EVIL-AI) projects by Jane and Aatos Erkko Foundation and the “Multifaceted ripple effects and limitations of human-AI interplay at work, business and society (SYNTHETICA)'' project (358714) by Research Council of Finland.
\end{acknowledgments}

\section*{Declaration on Generative AI}

During the preparation of this work, the authors used ChatGPT in order to perform grammar and spelling checks, and to enhance clarity in some portions of the text. After using this tool, the authors reviewed and edited the content as needed and take full responsibility for the publication’s content.

\bibliography{references}

\end{document}